\documentclass[runningheads]{llncs}
\usepackage{graphicx}
\usepackage{xcolor}
\usepackage{amsmath}
\usepackage{amssymb}
\usepackage{multirow}
\usepackage[figuresright]{rotating}
\usepackage{tabularx}
\makeatletter
\def\hlinewd#1{%
\noalign{\ifnum0=`}\fi\hrule \@height #1 %
\futurelet\reserved@a\@xhline}
\makeatother

\begin{document}
\title{SA-GAN: Structure-Aware GAN for Organ-Preserving Synthetic CT Generation}
\titlerunning{SA-GAN: Structure-Aware GAN}

\author{Hajar Emami\inst{1}%\orcidID{0000-1111-2222-3333} 
% index{Last Name, First Name}
\and
Ming Dong\inst{1}%\orcidID{1111-2222-3333-4444} 
% index{Last Name, First Name}
\and
Siamak P. Nejad-Davarani\inst{2}%\orcidID{2222--3333-4444-5555} 
% index{Last Name, First Name}
\and
Carri K. Glide-Hurst\inst{3}%\orcidID{2222--3333-4444-5555} 
% index{Last Name, First Name}
}
\authorrunning{H. Emami et al.}

\institute{Department of Computer Science, Wayne State University, MI 48202, USA 
\email{mdong@wayne.edu}\\
\and
Department of Radiation Oncology, University of Michigan, MI 48109, USA
\and
Department of Human Oncology, University of Wisconsin Madison, WI 53792, USA}
\maketitle              % typeset the header of the contribution
\begin{abstract}
In medical image synthesis, model training could be challenging due to the inconsistencies between images of different modalities even with the same patient, typically caused by internal status/tissue changes as different modalities are usually obtained at a different time. This paper proposes a novel deep learning method, Structure-aware Generative Adversarial Network (SA-GAN), that preserves the shapes and locations of in-consistent structures when generating medical images. SA-GAN is employed to generate synthetic computed tomography (synCT) images from magnetic resonance imaging (MRI) with two parallel streams: the global stream translates the input from the MRI to the CT domain while the local stream automatically segments the inconsistent organs, maintains their locations and shapes in MRI, and translates the organ intensities to CT. Through extensive experiments on a pelvic dataset, we demonstrate that SA-GAN provides clinically acceptable accuracy on both synCTs and organ segmentation and supports MR-only treatment planning in disease sites with internal organ status changes.

\keywords{Structure-aware GAN \and Synthetic CT \and Radiation Therapy.}
\end{abstract}
\section{Introduction}
\label{intro}
Multimodal medical imaging is crucial in clinical practice such as disease diagnosis and treatment planning. For example, Computed tomography (CT) imaging is an essential modality in treatment planning. Compared with CT, Magnetic Resonance Imaging (MRI) is a safer modality that does not involve patient's exposure to radiation. Due to the excellent soft tissue contrast in MRI, its integration into CT-based radiation therapy is expanding at a rapid pace. However, separate acquisition of multiple modalities is time-consuming, costly and increases unnecessary irradiation to patients. Thus, a strong clinical need exists to synthesize CT images from MRI \cite{fu2019deep,kim2015dosimetric,huynh2015estimating}.

\begin{figure}[htbp]
\centering
\includegraphics[width=0.58\columnwidth]{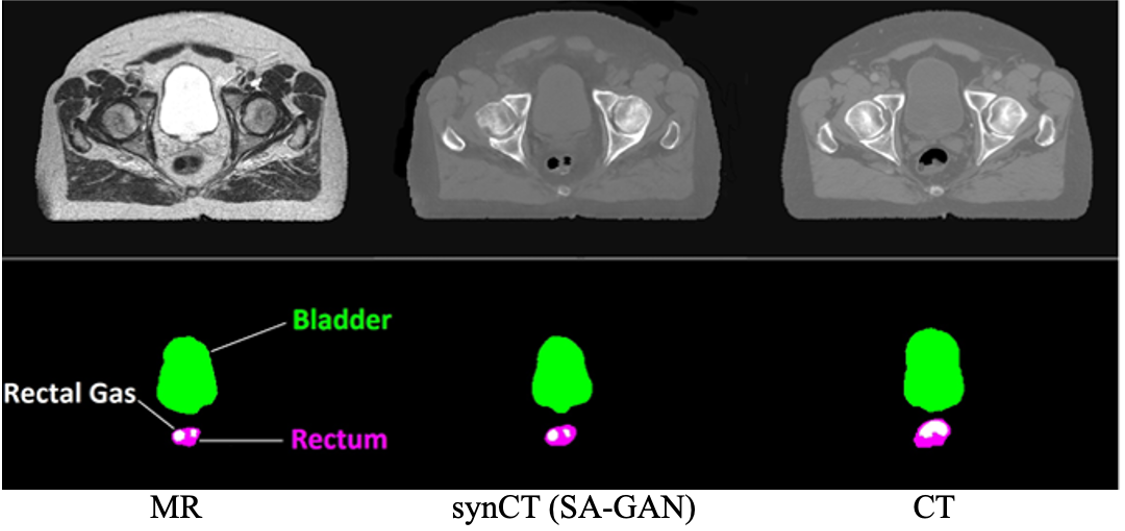}
\caption{Example of pelvic MRI with the corresponding CT, highlighting significant inconsistencies between the MRI and CT in bladder, rectum and rectal gas. The synCT generated by SA-GAN preserves the location and shape of organs as in MRI while changing their intensity to the CT, leading to more accurate dose calculation.}%in radiation therapy.}
\label{fig:example}
\end{figure}

Recently, deep learning methods have been employed in medical image synthesis. Initially, Fully Convolutional Networks (FCNs) were used in medical image synthesis \cite{nie2016estimating,han2017mr,emami2020frea}. The adversarial training in conditional generative adversarial networks (cGANs) \cite{isola2017image,goodfellow2014generative,lin2018conditional} can retain fine details in medical image synthesis and has further improved the performance over CNN models \cite{nie2017medical,emami2018generating,armanious2020medgan,emami2020attention}. Cycle generative adversarial network (CycleGAN) \cite{zhu2017unpaired} has also been used for unsupervised medical image translation \cite{lei2019mri,wolterink2017deep,hamghalam2020high,pan2018synthesizing} when paired images are not available. While deep learning techniques have made enormous achievement in medical image synthesis, model training could be challenging due to the inconsistencies between images of different modalities even with the same patient, typically caused by internal status/tissue changes as different modalities are usually obtained at a different time. For instance, in prostate cancer treatment, MRI and the corresponding CT slices of pelvis are often not consistent due to variations in bladder filling/emptying, irregular rectal movement, and local deformations of soft tissue organs \cite{nakamura2010variability}. 
Thus, it is of paramount importance to generate synCTs from MRI that can accurately depict these variations to spare critical organs while ensuring accurate dose calculation in MR-only treatment planning.

In Fig. \ref{fig:example}, an example of pelvic MRI is shown with the corresponding CT with a clear inconsistency in the rectal gas regions. Also note that the variations in bladder filling result in significant organ shape changes between MRI and the corresponding CT. This would adversely impact the dose delivered to the prostate as accurately preserving bladder status is important for radiation therapy. The desired synCT from the MRI example is shown in the second column, which should preserve the location and shapes of the inconsistent organs as in MRI while accurately changing the corresponding image intensities to the real CT to ensure high fidelity dose calculation in radiation therapy.

To improve MRI-CT consistency, pre-processing is typically employed. Chen et al. \cite{chen2018u} utilized deformable registration to handle the MRI-CT inconsistencies. However, the accuracy of synCTs will largely depend upon the performance of the deformable registration, which is limited in multi-modality workflows with large volume changes and introduces geometrical uncertainties in the pelvic region  \cite{zhong2015adaptive}. Maspero et al.\cite{maspero2018dose} assigned gas regions from MRI to CT as a pre-processing step to ensure the consistency of gas regions. This manual intervention is simple but time consuming. More importantly, it is restricted to a particular type of inconsistency (rectal gas in a pelvic), lacking the generality in the applicability to other types of inconsistencies or other disease sites. More recently, Ge et al.\cite{ge2019unpaired} used shape consistency loss in training to align the skin surface and overall shape of the synCT images with the original input image.

This work proposes a novel deep learning method, a structure-aware GAN (SA-GAN) model, that preserves the shapes and locations of inconsistent structures when generating synCTs. Our SA-GAN model is developed on the basis of GAN with two parallel streams in the generator. The global stream translates the input from the source domain (MRI) to the target domain (CT) while the local stream automatically segments the inconsistent organs between the two domains using a structure segmentation network, maintains their locations and shapes in MRI, and translates the organ intensities to CT using a novel adaptive organ normalization (AdaON) module. Our major contributions are: \textbf{1)} SA-GAN is the first automated framework that systematically addresses the MR-CT inconsistent issues in synCT generation in an end-to-end fashion. \textbf{2)} By fusing outputs from both the global and local streams, SA-GAN jointly minimizes the reconstruction and the structure segmentation losses together with the GAN loss through adversarial training. \textbf{3)} Without the time-consuming and error-prone pre-processing, SA-GAN offers strong potential for near real-time MR-only treatment planning while preserving features necessary for high precision dose calculation.

\section{Method}
The goal of the proposed SA-GAN model is to estimate a mapping $F_{MR \rightarrow CT}$ from the source domain (MRI) to the target domain (CT). The mapping F is learned from paired training data $S=\{(mr_i,ct_i)|mr_i\in{MR},ct_i\in{CT},i=1,2,...,N\}$, where N is the number of MRI-CT pairs. This translation is a challenging task as some regions are not completely matched between two domains, e.g., bladder, rectum and rectal gas regions in a pelvic dataset. A conventional synCT model typically defines the same loss function across all regions in an input MRI. This would produce erroneous tissue assignments in the inconsistent regions between two domains. To better preserve organ features and shapes in synCT generation, we need to solve two important tasks simultaneously: (1) segmenting the inconsistent areas and translating only the image intensities in these regions (2) translating the remaining regions paired between MRI and CT.

\begin{figure}[t]
\centering
   \includegraphics[width=1\linewidth]{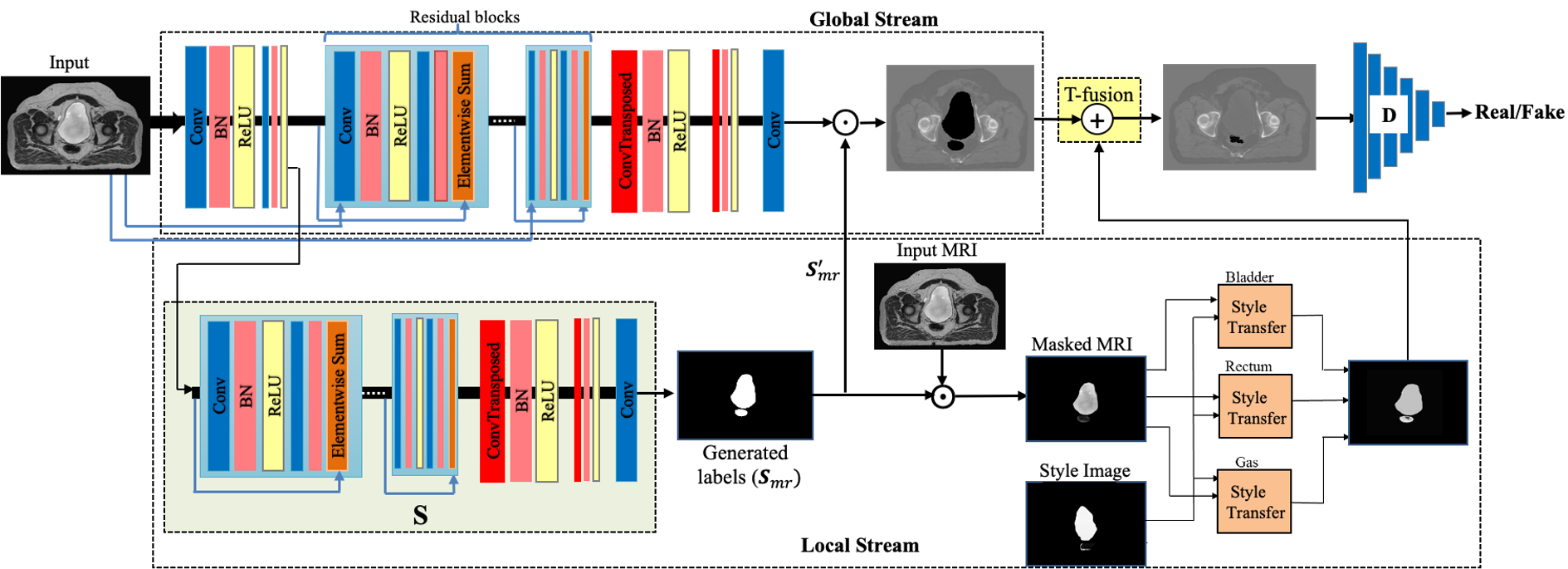}
   \caption{SA-GAN architecture with two parallel streams.}
   %\textcolor{red}{The global stream translates the input from the MRI to the CT domain while the local stream automatically segments the inconsistent organs between the two domains using a structure segmentation network (S), keeps their locations and shapes in MRI, and translates the image intensities to CT through the style transfer modules. The outputs of two streams are fused and fed into the discriminator D. Skip connections are shown by blue arrows.}}
   \label{fig:Fig1}
\end{figure}

The architecture of our proposed solution, SA-GAN, is shown in Fig.~\ref{fig:Fig1}. The main components include the generator ($G$), the discriminator ($D$), the structure segmentation network ($S$), the style transfer modules with adaptive organ normalization (AdaON) layers, and the fusion layer ($T_{fusion}$). SA-GAN contains two parallel streams in the generator. The global stream translates the input from MRI to CT domain while the local stream automatically segments the inconsistent regions using network $S$, keeps their locations and shapes in the MRI space, and translates the organ intensities to CT intensities using AdaON. The outputs of two streams are combined in the fusion layer to generate the final synCT. SA-GAN's global stream is used for global translation of consistent regions between the two domains and produces the reconstruction loss, and the local stream segments the inconsistent organs and produces the structure segmentation loss. These losses are jointly minimized together with the GAN loss through adversarial training in an end-to-end fashion.

\subsection{Global stream in SA-GAN} 
The generator $G$ includes an encoder-decoder network with three convolution layers followed by nine residual blocks. Each residual block consist of two convolutional layers followed by batch normalization and a nonlinear activation function: Rectified Linear Unit (ReLU). In addition to the skip connections in the residual blocks, we also add connections from the input layer to the layers of the encoder. The input is down-sampled and is concatenated with the feature maps of the encoder layers. Adding these connections can help the network to preserve the low level features of the input MRI such as edges and boundaries. Finally, we have three transposed convolutional layers after the residual blocks to generate synCTs of the input size. These transposed convolution layers are usually used in the decoder part of an encoder-decoder architecture to project feature maps to a higher dimensional space. The dropout units are also used in the layers of the generator for regularization and to help prevent overfitting.

The discriminator (D) is a CNN with six convolutional layers followed by batch normalization and ReLU. $G$ tries to generate synCTs as close as possible to real CTs while $D$ tries to distinguish between the generated synCTs and real CTs. Following \cite{mao2017least}, we replace the negative log likelihood objective by a least square loss to achieve more stable training:
\begin{align}
\mathcal{L}_{GAN}(G,D)=\mathbb{E}_{ct{\sim}P(ct)}[(D(ct)-1)^2]+\mathbb{E}_{mr{\sim}P(mr)}[D(G(mr))^2]
\end{align}
\noindent where $P$ is the data probability distribution. Moreover, $L_1$ norm is also used as the regularization in the reconstruction error:
\begin{align}
\mathcal{L}_{L_1}(G) =\mathbb{E}_{mr{\sim}P(mr),ct{\sim}P(ct)}[||G(mr)-ct||_1]
\end{align}
Using $L_1$ distance encourages less blurred results when compared with using $L_2$ to compute the reconstruction error \cite{isola2017image}. Because of the potential MRI-CT inconsistencies, the reconstruction loss should not be calculated for all regions of the input. If inconsistent regions are included when calculating $L_1$ reconstruction loss, the error could be agnostic and will possibly mislead the generator. Thus, we need to exclude these regions in our calculations. To do so, we first define a binary mask $u$ for the excluded  regions as the union of inconsistent regions within MRI ($label_{mr}$) and CT ($label_{ct}$) provided in the ground truth:
\begin{align}
u = label_{mr} \cup label_{ct}
\end{align}
where $u$ is 0 for regions to be excluded and 1 for all other regions. Then, the reconstruction loss is modified by performing element-wise multiplication on the real and the generated CT with $u$:
\begin{align}
\mathcal{L}_{exc}(G) =\mathbb{E}_{mr{\sim}P(mr),ct{\sim}P(ct)}[||u \odot G(mr)-u \odot ct||_1]
\end{align}
where $\odot$ denotes the element-wise multiplication operator. Note that MRI and CT labels of inconsistent regions are used here to calculate the modified reconstruction loss. After training, these labels are not required anymore for generating synCT. Finally, the objective is the minimax optimization defined as:
\begin{align}
G^*_{MR \rightarrow CT}, D^* = arg \underset{ G} min \underset{ D} max \mathcal{L}_{GAN}(G_{MR \rightarrow CT},D)+ \lambda \mathcal{L}_{exc}(G)
\end{align}
where we set the loss hyper-parameter $\lambda$ = 10 throughout our experiments.

\subsection{Segmentation network in the local stream}
The segmentation network in SA-GAN takes an input MRI image and segments out the pre-defined inconsistent structures. The network $S$ has three convolution layers at the beginning, followed by six residual blocks and three transposed convolutional layers. The loss function to update $S$ is a multi-class cross-entropy:
\begin{align}
\mathcal{L}_{S} = \mathbb{E}_{mr{\sim}P(mr)} [-\frac{1}{N} \sum_{i} w_i^{-1} y^{i} log(S(mr)_i)]
\end{align}
where $y$ is the ground truth labels, and N is the total number of pixels. In order to balance the size differences of structures and their contribution to the loss, a coefficient $w_i$ is adopted to weight each structure $i$ with the invert sampling count: $w_i =\frac{n_i}{N}$ where $n_i$ is the number of pixels belonging to structure $i$.

\subsection{Organ style transfer with AdaON}
Research shows that the image style and content are separable, and it is possible to change the style of an image while preserving its content \cite{gatys2016image,ghiasi2017exploring}. In adaptive instance normalization (AdaIN) \cite{huang2017arbitrary}, an encoder-decoder is used to take a content image and a style image, and synthesizes an output image. After encoding the content and style images in feature space using encoder $f$, the mean and the standard deviation of the content input feature maps are aligned with those of the style input in an AdaIN layer. Then a randomly initialized decoder $g$ is trained to generate the stylized image from the output of the AdaIN layer. 

In SA-GAN, we use adaptive organ normalization (AdaON) layers to transfer the style of inconsistent organs to CT domain. To generate accurate results, we define a style for each inconsistent structure in the pelvic. Then, $AdaON_{B}$, $AdaON_{R}$ and $AdaON_{G}$ layers are trained separately for bladder, rectum and rectal gas regions by taking a masked CT style image and a masked MRI from an inconsistent region as inputs and minimizing style and content losses. After training, they are used to generate the inconsistent organs in the CT domain, which are then combined through element-wise addition as the output of the local stream. Finally, the outputs of the global and local streams are combined in the fusion layer $T$, also through element-wise addition.

\section{Experiments and Results}
\noindent\textbf{Dataset.} In our experiments, 3,375 2D images from 25 subjects
with prostate cancer (median age = 71 years, Range: 53-96 years) were retrospectively analyzed.
%as part of an IRB approved study. 
CT images were acquired on a Brilliance Big Bore scanner with 512 $\times$ 512 in-plane image dimensions, 1.28 $\times$ 1.28 $mm^2$ in-plane spatial resolution, and 3 mm slice thickness. Corresponding MR images were acquired in the radiation therapy treatment position on a 1.0 Tesla Panorama High Field Open MR simulator (576$\times$576$\times$250 $mm^3$, Voxel size = 0.65$\times$0.65$\times$2.5 $mm^3$). A single physician delineated bladder and rectum volumes on CT and MRI images in the Eclipse Treatment Planning System. Rectal gas was identified by thresholding and applying morphological operators. Next, using Statistical Parametric Mapping (SPM) 12, MRIs and corresponding binary masks of each subject were rigidly co-registered to the CT images of the same subject. 

The size of our dataset is consistent with the ones used for synCT in the literature \cite{nie2016estimating,nie2017medical,nie2018medical,han2017mr,emami2018generating,hamghalam2020high,lei2019mri,fu2019deep}. Additionally, we used data augmentation (e.g., flipping) to increase the number of training images four times. To evaluate our model performance and avoid overfitting, a five-fold cross-validation was used in our model training and testing.

\noindent\textbf{Implementation details.}
The weights in SA-GAN were all initialized from a Gaussian distribution with parameter values of 0 and 0.02 for mean and standard  deviation, respectively. The model is trained with ADAM optimizer with an initial learning rate of 0.0002 and with a batch size of 1. We trained the model for 200 epochs on a NVIDIA GTX 1080 Ti GPU. SA-GAN training took approximately 13 hours. The model was implemented using PyTorch. 
 
\noindent\textbf{Evaluation metrics.}
Three commonly-used quantitative measures are adopted for evaluation: Mean Absolute Error (MAE), Peak Signal-to-Noise Ratio (PSNR), and Structure Similarity Index (SSIM). For PSNR and SSIM, higher values indicate better results. For MAE, the lower the value, the better is the generation. Consideration was given to agreement using the full field of view, bone, rectal gas, bladder and rectum regions. Since input MRIs and their corresponding CTs include inconsistent regions in bladder, rectum and rectal gas, we calculate the MAE between the intersected regions in real CTs and synCTs. To segment the bone regions in the images, a threshold of +150 HU \cite{fu2018male} was set in both real and generated CTs. We used, Dice similarity coefficient (DSC), the well-established metric to evaluate the organ segmentation performance.

\noindent\textbf{Ablation study.}
We performed the following ablation study. 1) We removed the local stream in SA-GAN. In this case, our model is reduced to cGAN. 2) We removed the local stream but added style loss and content loss calculated from the whole image to the model's objective (SA-GAN-wo-$seg$). 3) We had the segmentation network in the local stream but removed the AdaON layers (SA-GAN-wo-$AdaON$). 4) We had the whole local stream but the reconstruction loss is calculated from all regions in an MRI input image (SA-GAN-wo-$\mathcal{L}_{exc}$). MAE, PSNR and SSIM for different configuration of our model are reported in the supplementary materials. In summary, SA-GAN with all its components achieved the best results in our model ablation.

\noindent\textbf{Comparison to State-of-the-Arts.} 
Extensive experiments were performed to compare SA-GAN with current state-of-the-art models for synCT generation. Since SA-GAN's generator and discriminator share a similar architecture to cGAN \cite{isola2017image,tie2020pseudo} and CycleGAN \cite{zhu2017unpaired}, we included these methods in our comparison. In particular, CycleGAN is widely used for medical image synthesis and provides a potential solution to handle the organ inconsistencies in the pelvic dataset as it does not require paired input images. It is worth mentioning that because of the flexible architecture of SA-GAN, other variants of cGAN and CycleGAN, e.g., \cite{armanious2020medgan} \cite{hamghalam2020high}, can easily be incorporated as the generator in the global stream of SA-GAN. All the comparisons are done using the same training and testing splits.

\begin{figure*}[t]
\centering
    \includegraphics[width=\textwidth]{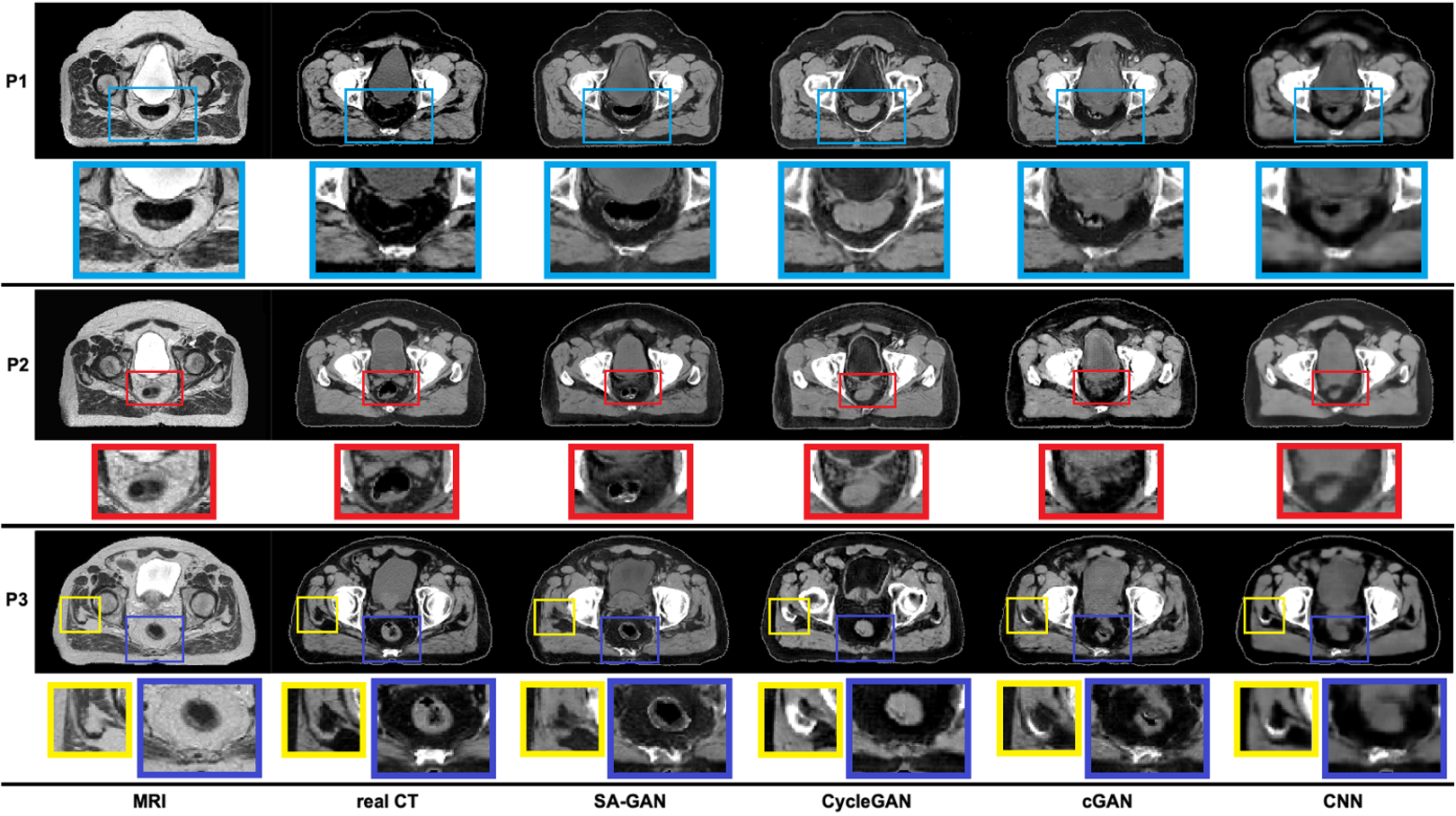}
   \caption{Qualitative comparison of synCTs generated by SA-GAN and state-of-the-art models for three different patients (P1-P3). Regions of interest are drawn on a slice and enlarged in the row below for a more detailed comparison with other models.}
\label{fig:comparison}
\end{figure*}

\noindent\textbf{SynCT generation results.}
Fig.~\ref{fig:comparison} shows synCT results for three different test cases that have inconsistencies in the bladder, rectum, and rectal gas between MRI and CT.
The CNN model with no discriminator block generated blurry results. Since the discriminator in cGAN, CycleGAN, and SA-GAN can easily classify blurry images as fake, they do not generate blurry results. However, cGAN and CycleGAN are not able to preserve the inconsistent regions, particularly for the transient rectal gas regions which has been shown to cause dose calculation discrepancies \cite{kim2015dosimetric}. As shown in Fig.~\ref{fig:comparison}, for Patient 3, substantial rectal gas appeared in the MRI that did not correspond to the CT. Consequently, CNN, cGAN, and CycleGAN yielded inaccurate intensities (similar to tissues) in this region, whereas rectal gas was accurately depicted in the results of SA-GAN. Similarly, Patients 1 and 2 had less air in their rectal regions in MRI than their corresponding CTs, and these were apparent in the SA-GAN results whereas all other methods failed to reproduce the appropriate tissue type and shape. SynCTs generated by the CNN, cGAN, and CycleGAN models showed larger errors in the bone regions. Bones were generated in non-physical regions, and their intensities were underestimated in the spinal region. Clearly, bone anatomy was better represented by SA-GAN. In general, our qualitative comparison shows that SA-GAN produces more realstic synCTs with a higher spatial resolution and preserved organ shapes in MRI when compared to other methods. Please see the supplementary material for more synCT examples with a bigger display.

The average MAE, PSNR, and SSIM computed based on the real and synthetic CTs for all test cases are listed in Table 1. To better evaluate the performance, the MAEs for different regions are also reported. SA-GAN achieved better performance in generating both consistent (e.g., bone) and inconsistent (e.g., rectal gas) regions when compared to other models over all test cases. Clearly, the improvement in the inconsistent regions is due to the structure segmentation and AdaON in the local stream of SA-GAN. 
Additionally, in the global stream, we compute the reconstruction error between synCTs and real CTs after excluding the inconsistent regions. This allows our global translation model to focus on the consistent regions when calculating the loss, leading to more accurate results than other models. Finally, it is worth mentioning that \textbf{the combination of consistent and inconsistent regions in MRI-CT pairs presents a significant challenge} for synCT methods based on pixel correspondences, e.g., cGAN and CNN-based methods. Although CycleGAN was first introduced to handle inconsistencies between two domains in the absence of paired data, as shown in our experimental results, it was \textbf{unable} to maintain the appropriate consistency between the bladder and rectal intensity values.

\begin{table*}[t]
\centering
\label{table:synCT}
\caption{Performance comparison of SA-GAN for synCT generation. The average MAEs are computed from entire pelvis, bone, rectal gas, rectum and bladder. Since MRIs and CTs include inconsistent regions, the MAE is calculated between the intersected masked regions of MRIs and real CTs in bladder, rectum and rectal gas.}
\resizebox{\textwidth}{!}{%
\begin{tabular}{c|c|c|c|c|c|c|c}
\hlinewd{1pt}
\multirow{2}{*}{Method} & \multicolumn{5}{c|}{MAE (HU)} & {\multirow{2}{*}{PSNR}} & \multirow{2}{*}{SSIM}\\
\cline{2-6}
%\hline
 & Entire pelvis & Bone & Rectal Gas & Rectum & Bladder & & \\
\hline
CNN&42.7 $\pm$ 4.9&239.4 $\pm$ 35.6&676.4 $\pm$ 85.7&28.3 $\pm$ 8.4&13.9 $\pm$ 2.1&29.9 $\pm$ 1.2&0.88 $\pm$ 0.05\\
\hline
cGAN&54.6 $\pm$ 6.8&267.2 $\pm$ 43.6&652.6 $\pm$ 97.4&35.7 $\pm$ 11.0&20.2 $\pm$ 7.8&28.0 $\pm$ 0.9&0.85 $\pm$ 0.04\\
\hline
CycleGAN&59.8 $\pm$ 6.1&290.3 $\pm$ 43.2&678.8 $\pm$ 75.6&35.9 $\pm$ 7.1&41.1 $\pm$ 16.6&27.2 $\pm$ 1.0&0.83 $\pm$ 0.05\\
\hline
SA-GAN&\textbf{38.5 $\pm$ 4.9}&\textbf{210.6 $\pm$ 34.0}&\textbf{279.9 $\pm$ 64.1}&\textbf{28.2 $\pm$ 7.2}&\textbf{13.5 $\pm$ 2.6}&\textbf{30.5 $\pm$ 1.3}&\textbf{0.90 $\pm$ 0.04}\\
\hlinewd{1pt}
\end{tabular}
}
\end{table*}

\begin{table}[t]
\centering
\label{table:Seg}
\caption{Dice similarity coefficient (higher is better) computed between the ground truth MRI labels and the segmented organs using SA-GAN's segmentation network.}
\begin{tabular}{c|c|c|c}
\hlinewd{1pt}
 & Bladder & Rectal Gas  & Rectum  \\
\hline
{} DSC {} & {} 0.93 $\pm$ 0.03 {} & {} 0.90 $\pm$ 0.05 {} & {} 0.86 $\pm$ 0.03 {}\\
\hlinewd{1pt}
\end{tabular}
\end{table}

\noindent\textbf{Organ segmentation results.}
The DSCs between the ground truth MRI labels and the segmented organs using SA-GAN are summarized in Table 2. High DSC suggests excellent ability of SA-GAN to localize inconsistent organs. Some segmentation examples are provided in the supplementary material.

Once trained, SA-GAN generates both the synCTs and segmentation labels of the inconsistent regions in the input MRIs in a short time of $\sim$12 s. Thus, SA-GAN provides a practical tool that can be integrated into near real-time applications for producing synCTs in datasets with inconsistent structures.

\section{Conclusion}
In this paper, we proposed a novel SA-GAN to generate synCTs from MRI for disease sites where internal anatomy may change between different image acquisitions. Our experiments show that SA-GAN achieves clinically acceptable accuracy on both synCTs and organ-at-risk segmentation, and thus supports MR-only treatment planning (e.g., high fidelity dose calculation in radiation therapy) in disease sites with internal organ status changes.

\noindent\textbf{Acknowledgments.} This work was partially supported by the National Cancer Institute of the National Institutes of Health under Award Number R01CA204189.

\bibliographystyle{splncs04}
\bibliography{mybibliography}

\clearpage

\noindent\textbf{Supplementary Material}

\begin{table}
\centering
\label{table:Ablation}
\caption{Ablation study of key components in SA-GAN.}
\begin{tabular}{c|c|c|c}
\hlinewd{1pt}
Method & MAE & PSNR  & SSIM\\
\hline
cGAN&54.6 $\pm$ 6.8&28.0&0.85\\
%\hline
SA-GAN-wo-$seg$&51.4$\pm$7.9&28.4&0.86\\
%\hline
SA-GAN-wo-$AdaON$&49.5$\pm$6.2&28.5&0.86\\
%\hline
SA-GAN-wo-$\mathcal{L}_{exc}$&42.6$\pm$6.8&30.2&0.88\\
%\hline
SA-GAN&\textbf{38.5 $\pm$ 4.9}&\textbf{30.5}&\textbf{0.90}\\
\hlinewd{1pt}
\end{tabular}
      \vspace{-0.4cm}
\end{table}

\begin{figure}
\centering
   \includegraphics[width=0.8\linewidth]{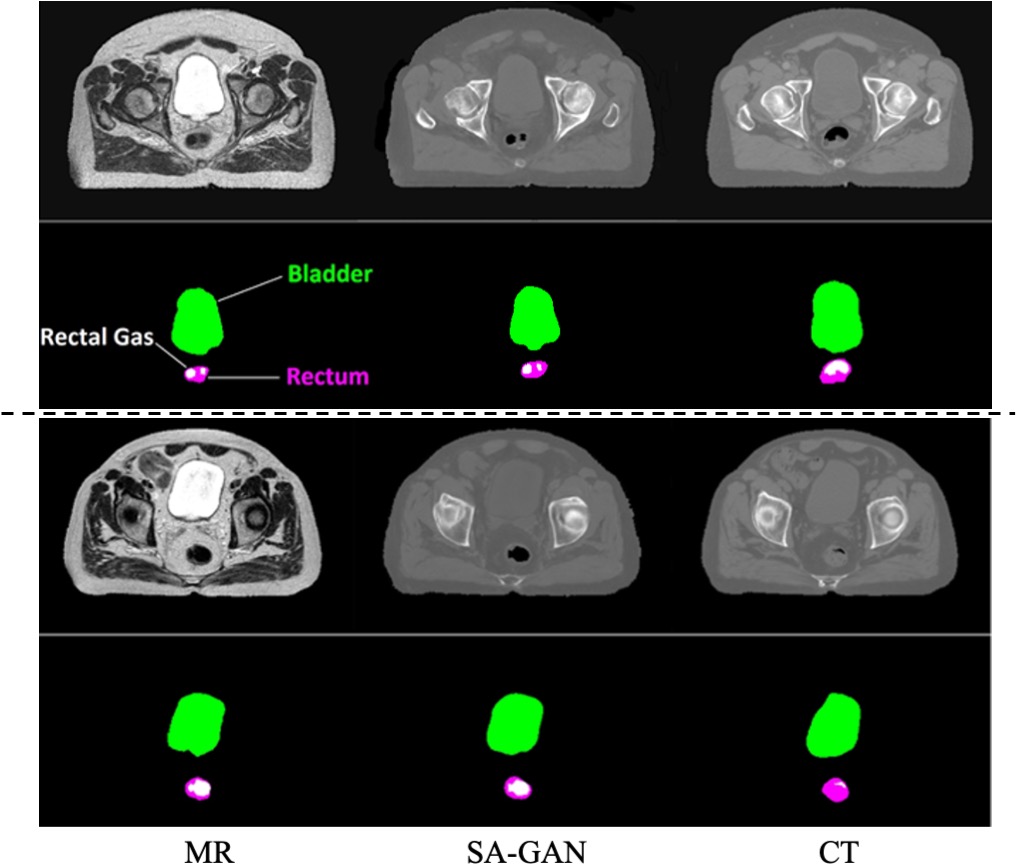}
      \vspace{-0.2cm}
   \caption{Examples of segmentation results for two different prostate cancer test cases that have inconsistent organs/gas in MRI and CT. The ground truth of input MRI organs/gas, real CT organs/gas, and segmented organs/gas in the local stream are shown in the second row. Bladder, rectal gas and rectum are
shown by green, white and pink, respectively. }
\label{fig:seg}
      \vspace{-0.1cm}
\end{figure}

\begin{figure}[htbp]
    \centering
    \includegraphics[width=\textwidth]{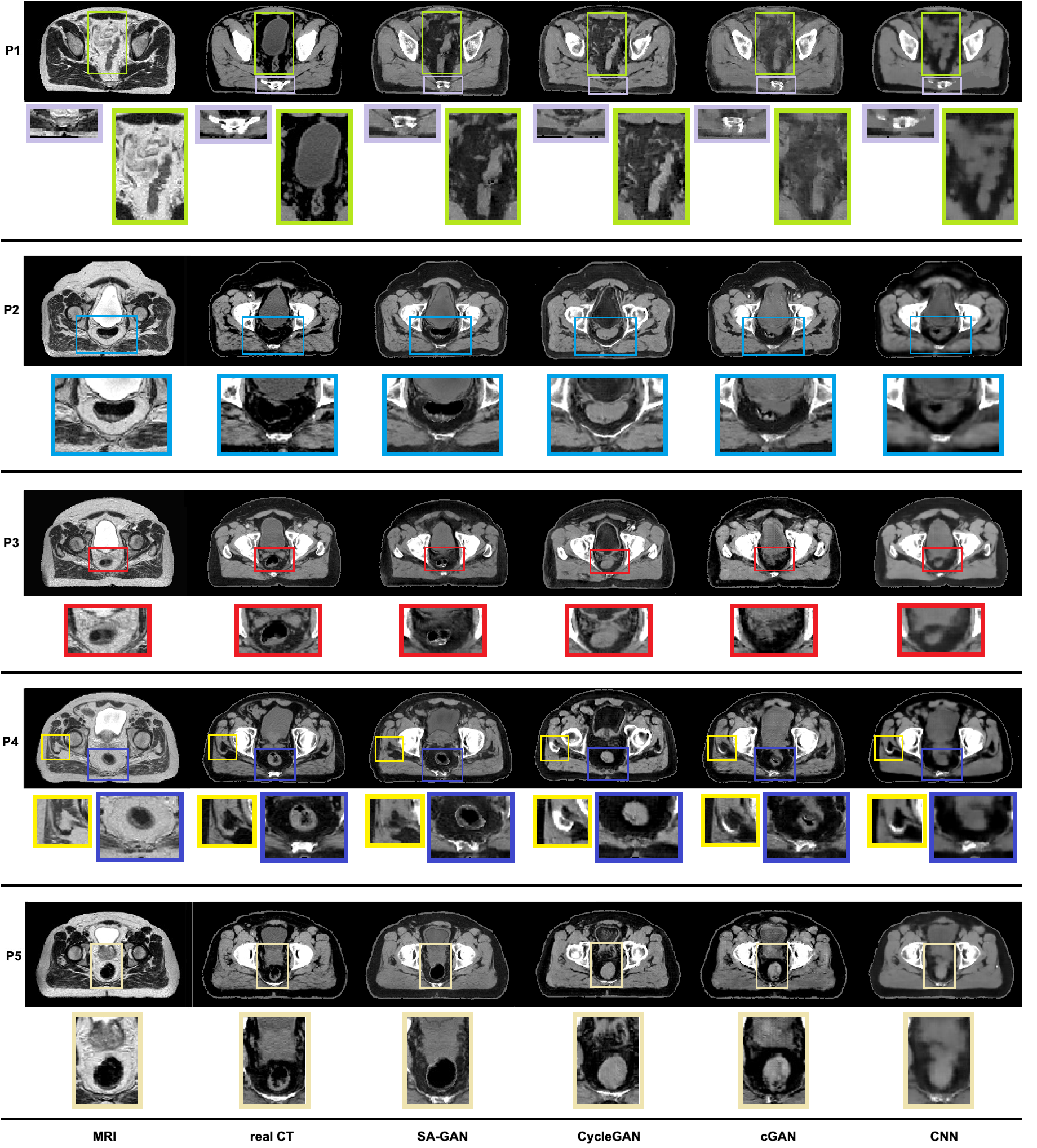}
   \caption{Qualitative comparison of synCTs generated by SA-GAN and state-of-the-art models for five different patients (P1-P5). The input MRI, corresponding real CT and synCT generated by SA-GAN, CycleGAN, cGAN and CNN, are shown in the columns, respectively for five different patients with prostate cancer. Regions of interest are drawn on a slice and enlarged in the row below for a more detailed comparison with other models.
   }
\label{fig:comparison2}
\end{figure}

\end{document}